\documentclass[aps,prb,preprint,amsmath,amssymb,superscriptaddress]{revtex4}
\usepackage{graphicx}
\usepackage{dcolumn}
\usepackage{bm}
\usepackage{amsmath}
\usepackage{amssymb}
\begin{document}
\title{Spin Waves and Switching: The Dynamics of Exchange - Biased
Co Core - CoO Shell Nanoparticles}
\author{Mikhail Feygenson}
\affiliation{Condensed Matter Physics and Materials Science Department, Brookhaven National Laboratory, Upton, New York 11973, USA}
\author{Xiaowei Teng}
\affiliation{Center for Functional Nanomaterials, Brookhaven National Laboratory, Upton, New York 11973, USA}
\affiliation{Department of Chemical Engineering, University of New Hampshire, Durham, NH 03824, USA}
\author{Sue E. Inderhees}
\affiliation{Department of Physics, University of Michigan, Ann Arbor MI 48109-1120, USA}
\author{Yuen Yiu}
\affiliation{Condensed Matter Physics and Materials Science Department, Brookhaven National Laboratory, Upton, New York 11973, USA} \affiliation{Department of Physics and Astronomy, Stony Brook University, Stony Brook NY 11794, USA}
\author{Wenxin Du}
\affiliation{Department of Chemical Engineering, University of New Hampshire, Durham, NH 03824, USA}
\author{Weiqiang Han}
\affiliation{Center for Functional Nanomaterials, Brookhaven National Laboratory, Upton, New York 11973, USA}
\author{Jinsheng Wen}
\affiliation{Condensed Matter Physics and Materials Science Department, Brookhaven National Laboratory, Upton, New York 11973, USA}
\author{Zhijung Xu}
\affiliation{Condensed Matter Physics and Materials Science Department, Brookhaven National Laboratory, Upton, New York 11973, USA}
\author{Andrey A. Podlesnyak }
\affiliation{Spallation Neutron Source, Oak Ridge National Laboratory, Oak Ridge, TN 37831, USA}
\author{Jennifer L. Niedziela}
\affiliation{Spallation Neutron Source, Oak Ridge National Laboratory, Oak Ridge, TN 37831, USA}
\author{Mark Hagen}
\affiliation{Spallation Neutron Source, Oak Ridge National Laboratory, Oak Ridge, TN 37831, USA}
\author{Yiming Qiu}
\affiliation{NIST Center for Neutron Research, National Institute of Standards and Technology, 100 Bureau Dr., MS 6102, Gaithersburg, MD 20899, USA} \affiliation{Department of Materials Science and Engineering, University of Maryland, College Park, MD  20742, USA}
\author{Craig M. Brown}
\affiliation{NIST Center for Neutron Research, National Institute of Standards and Technology, 100 Bureau Dr., MS 6102, Gaithersburg, MD 20899, USA}
\author{Lihua Zhang}
\affiliation{Center for Functional Nanomaterials, Brookhaven National Laboratory, Upton, New York 11973, USA}
\author{Meigan C. Aronson}
\affiliation{Condensed Matter Physics and Materials Science Department, Brookhaven National Laboratory, Upton, New York 11973, USA} \affiliation{Department of Physics and Astronomy, Stony Brook University, Stony Brook NY 11794, USA}
\date{\today}
\begin{abstract}
The utility of nanoscaled ferromagnetic particles requires both stabilized moments and maximized switching speeds. During reversal, the spatial modulation of the nanoparticle magnetization evolves in time, and the energy differences between each new configuration are accomodated by the absorption or emission spin waves with different wavelengths and energy profiles. The switching speed is limited by how quickly this spin wave energy is dissipated. We present here the first observation of dispersing spin waves in a nanoscaled system, using neutron scattering to detect spin waves in the CoO shells of exchange biased Co core- CoO shell nanoparticles. Their dispersion is little affected by finite size effects, but the spectral weight shifts to energies and wave vectors which increase with decreasing system size. Core-shell coupling leads to a substantial enhancement of the CoO spin wave population above its conventional thermal level, suggesting a new mechanism for dissipating core switching energy.
\end{abstract}
\maketitle 
Due to intense experimental and theoretical effort, much is known about the structure and dynamics of nanoscaled systems where coupling between ferromagnet and antiferromagnet leads to the exchange bias effect.~\cite{nogues1999,stamps2000,nogues2005} In the exchange-biased state found below the blocking temperature T$_{B}$,  superparamagnetic reversals of the ferromagnetic moment become increasingly unlikely. Here, the dominant magnetic mode is the uniform precession of the core moments around an effective field H$_{eff}$,~\cite{cehovin2003,morup2005,morup2007} which is determined by both the magnetocrystalline and shape anisotropies, as well as the exchange field from the antiferromagnet. This precessional mode is the zero wave vector spin wave, and it has been detected in a number of different experiments,~\cite{stoecklein1988,mathieu1998,ercole2000} most notably in inelastic neutron scattering measurements where it appears as an inelastic peak with a characteristic energy in the range 0.01 - 0.1 meV.~\cite{hennion1994,hansen2000,lefmann2001,klausen2004,morup2005,morup2007} The direction of core precession can be reversed by the application of a magnetic field, and there is increasing evidence~\cite{safonov2004} that once the precession angle exceeds a critical value there is a sudden jump of the precession angle accompanied by an increase in dissipation, explained as an instability where the precessional mode decays into spin waves with a range of energies and wave vectors.  ~\cite{safonov2000,visscher2002,silva2002,safonov2004,livesey2007,wegelin2007} Experiments have so far not provided theory with the essential input which is needed to understand switching:  an experimentally determined spin wave spectrum in an exchange-biased system. Since exchange-biased systems are necessarily nanoscaled,  these spin wave spectra are expected to be significantly altered from their bulk counterparts both by finite size effects~\cite{hendriksen1993,wesselinowa2007} and by the exchange coupling between the antiferromagnetic and ferromagnetic components.~\cite{stamps1997} We stress as well that knowledge of the spin wave dispersion and density of states is essential for assessing the impact of dynamics on the equilibrium properties of nanoscaled systems, such as the uniform magnetization,~\cite{morup2005} and the electrical and thermal conductivities. It is increasingly problematic that so little is known about the magnetic dynamics of nanoscaled systems on length scales which are much smaller than the system size, and on time scales which are short compared to the switching times. We address this need by reporting here the first observation of dispersing spin waves in a nanoscaled system, exchange biased Co core/CoO shell nanoparticles.

We present here the results of a neutron scattering investigation of spin wave dynamics in ferromagnetic Co core - antiferromagnetic CoO shell nanoparticles where highly ordered core-shell interfaces and rigorous control of the core and shell dimensions (Table 1) enable a strong exchange bias effect.~\cite{inderhees2008} Our measurements show that the dynamics of the ferromagnetic core are increasingly affected as the temperature drops below the N\'{e}el temperature T$_{N}$ of the antiferromagnetic shell, and that the slowing of the core reorientation generates excess paramagnetic scattering in the CoO shell. Below the blocking temperature T$_{B}$ where core reorientation is energetically unfavorable, we find that the precession of the core moments is damped by the generation of inelastic spin waves in the antiferromagnetic shell. While previous neutron scattering measurements have documented both the superparamagnetic behavior of bare ferromagnetic and antiferromagnetic particles above the blocking temperature and their uniform precession at lower temperatures,~\cite{hansen2000,hansen1997,klausen2003} our measurements are the first to use neutron scattering techniques to study the coupling of magnetic dynamics in the ferromagnetic core and antiferromagnetic shell, and the first in any magnetic nanosystem to directly identify spin waves with nonzero wave vector q. In our experiments the reversal of the ferromagnetic cores occurs spontaneously in thermal equilibrium and in zero field, and in this way our results are complementary to those of ultrafast pump-probe experiments where nonequilibrium magnetization dynamics are investigated.~\cite{beaurepaire1996,ju1999,vankampen2002,weber2004,kimel2009} Our measurements find that exchange coupling introduces a new channel for energy dissipation which has not previously been investigated, namely the excitation of auxiliary spin waves in the antiferromagnetic shell, and suggest that magnetic switching times in systems with strong exchange coupling may be substantially enhanced relative to those of bare ferromagnetic particles.

\begin{table}
\caption{The average dimensions in nanometers of the nanoparticle powders used in the neutron scattering measurements. Error bars are as indicated.  R$_{tot}$ and R$_{core}$ are the radii of the nanoparticle and the Co core, respectively, and t$_{shell}$ is the thickness of the CoO shell.  H$_{EB}$ is the maximum exchange bias field observed at 30 K,  while \# is the total number of particles present in the neutron scattering powders. Details of the synthesis and characterization procedures used to establish these quantities are presented in the supplementary materials.\label{samples}}
\begin{ruledtabular}
\begin{tabular}{cccccc}
sample & R$_{tot}$ & R$_{core}$ & t$_{shell}$ & H$_{EB}$(Oe) & \# $\times$ $10^{17}$ \\
\hline
Co/CoO\#1 & 3.6(0.4) & 1.9(0.3) & 1.7(0.3) & 1500 & 5.6\\
Co/CoO\#2 & 4.5(0.5) & 2.5(0.4) & 2.0(0.3) & 6000 & 4.4 \\
Co/CoO\#3 & 5.5(0.5) & 1.4(0.2) & 4.1(0.5) & 814 & 3.6\\
Co/CoO\#4 & 5.5(0.5) & 1.0(0.2) & 4.5(0.5) & 435 & 4.8 \\
Co/CoO\#5 & 4.2(0.5) & 1.2(0.2) & 3.0(0.5)& 106 & 2.6 \\
CoO nps & 50(10) & 0 & 50(10) & 0 & 0.02\\
\end{tabular}
\end{ruledtabular}
\end{table}

\begin{figure*}
\includegraphics[width=15cm]{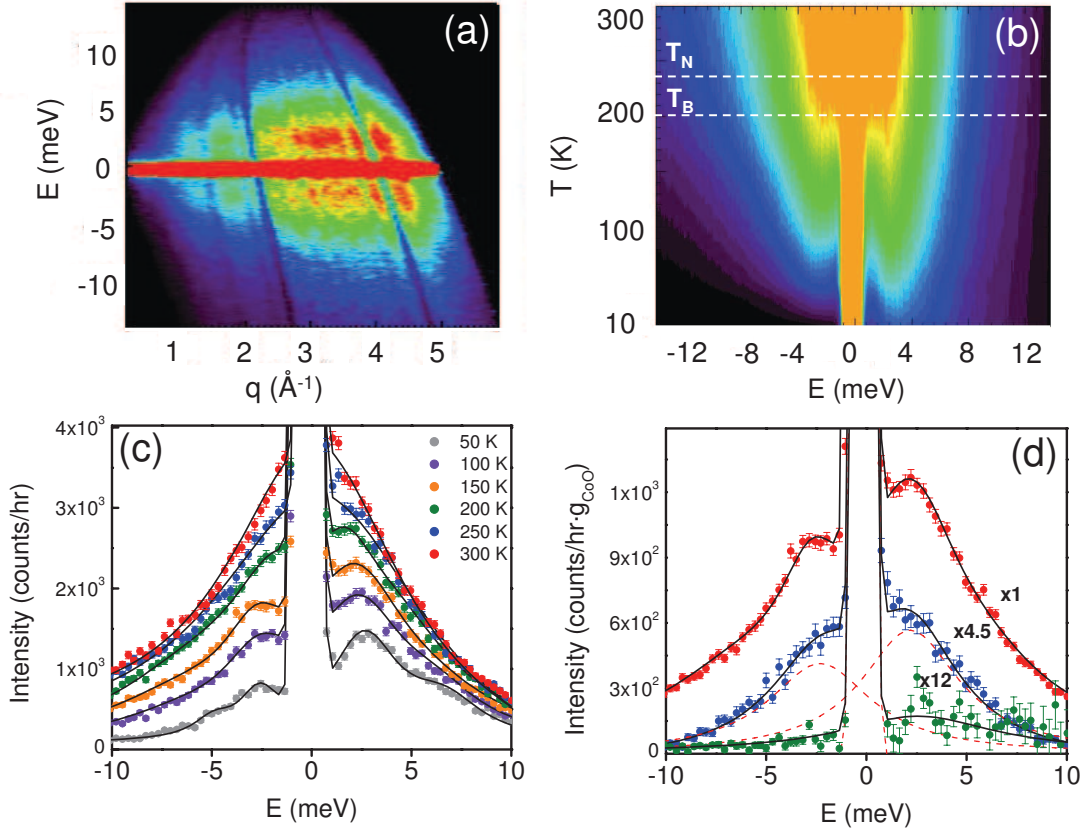}
\caption{\textbf{Excitations of Co Core/CoO Shell Nanoparticles.} We carried out zero field inelastic neutron scattering experiments on nanoparticle powders Co/CoO$\#$1-4 at the NIST Center for Neutron Research with the Disk Chopper Spectrometer (DCS), and used  the Cold Neutron Chopper Spectrometer (CNCS) at the Spallation Neutron Source to perform similar measurements on Co/CoO$\#$4/5, as well as two samples of pure CoO nanoparticles taken from different but nominally identical preparation batches, and finally a powder made by crushing a 5.3 g single crystal of CoO. (a) Contour plot of the scattered intensity I(q,E) for core-shell sample Co/CoO$\#$4 at 150 K, as a function of neutron energy transfer E and wave vector q. Dark lines are an experimental artifact originating with the radial collimator of CNCS. (b) The evolution of the q-integrated scattered intensity I(E) with temperature from 300 K - 10 K for sample Co/CoO$\#$4. Dashed lines indicate the blocking and N\'{e}el temperatures T$_B$ and T$_N$, respectively. (c) I(q,E) for a constant wave vector cut 0.28 $\AA^{-1}$ wide and centered at q = 1.40 $\AA^{-1}$, showing the evolution from quasielastic to inelastic excitations with decreasing temperature, as indicated. (d) Comparison of I(q,E) at 150 K and the same wave vector cut as (c) for Co/CoO$\#$4 (red points), fully oxidized CoO nanoparticles (blue points), and a powder of bulk CoO (green points), each normalized by the CoO mass. Solid lines are fits to functions described in the Supplementary Materials. Error bars in all figures represent 1$\sigma$.\label{overview}}
\end{figure*}

A contour plot of the scattered intensity I(q,E) for sample Co/CoO$\#$4 at 150 K as a function of neutron energy transfer E and wave vector q is presented in Fig.~\ref{overview}a. I(q,E) consists primarily of a pronounced inelastic peak, centered near $\approx$ 2 meV, which becomes more intense with increasing q. The temperature evolution of the q-integrated intensity I(E) is presented in Fig.~\ref{overview}b, and since T$_{N}$ and T$_{B}$ are the same for all five samples, the dynamics in each core-shell sample develops in a similar way. Above the N\'{e}el temperature of the CoO shell T$_N$ = 235 K,  the scattering is broad and quasielastic, with an intensity which diminishes as the temperature approaches the blocking temperature T$_B$ = 200 K. The quasielastic scattering collapses and an inelastic excitation centered at $\approx$2 meV emerges below T$_B$, shifting to larger energies and becoming more intense as the temperature is lowered. We will next argue that the quasielastic scattering found for T$\geq$ T$_{N}$ is due to a combination of paramagnetic scattering from the CoO shell and superparamagnetic fluctuations of the Co core, and that their disappearance with the ordering of the CoO shell and the onset of the exchange bias effect coincides with the onset of inelastic scattering, which we ascribe to conventional spin waves in the CoO shell.

\begin{figure*}
\includegraphics[width=18cm]{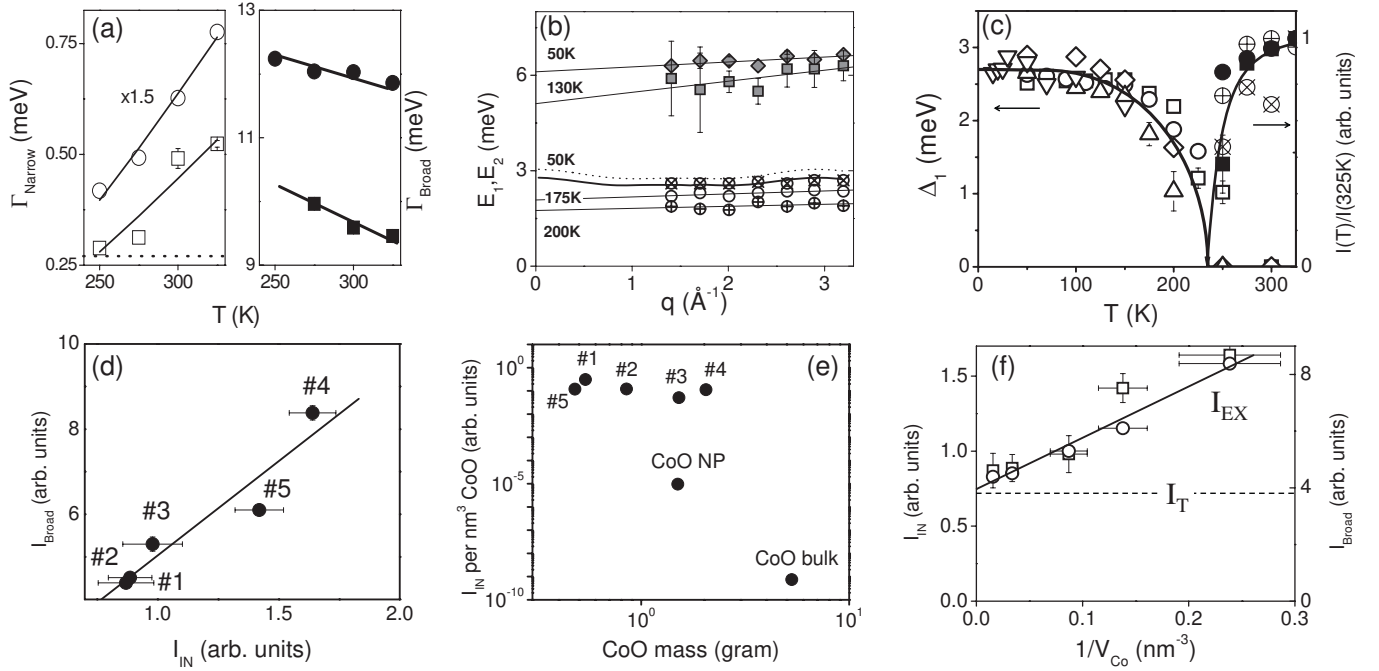}
\caption{\textbf{Superparamagnetism and the spin waves of nanoparticles}. (a) Left panel:  the temperature dependencies of the core reversal linewidths $\Gamma_{Narrow}$ ($\square$=Co/CoO$\#$3, $\circ$=Co/CoO$\#$4). Solid lines are fits to an Arrhenius expression $\Gamma_{Narrow}$(T)=$\Gamma_{N,0}$exp(-KV$_{Co}$/k$_{B}$T) and the horizontal dotted line shows the DCS experimental energy resolution of 0.27 meV. Right panel: temperature dependencies of the linewidths for paramagnetic fluctuations of the CoO shells $\Gamma_{Broad}$ ($\blacksquare$=Co/CoO$\#$3, $\bullet$=Co/CoO$\#4$). Solid lines are guides for the eye. (b) Dispersions of the inelastic peak center energies E$_{1}$(q)(open symbols) and E$_{2}$(q) (gray filled symbols) of Co/CoO$\#$4 at different temperatures, where the straight lines define the q=0 gaps $\Delta_{1}$(T) and $\Delta_{2}$(T). Dotted line shows spin wave dispersion calculated from bulk CoO parameters cited in the literature.~\cite{sakurai1968,daniel1969}. Dark solid line is the best fit of this model to our measured spin wave dispersion E$_{1}$(q). Relative to bulk CoO, there is no detectable increase in the first and second neighbor exchange couplings J$_{1}$ and J$_{2}$, only a small reduction in $\Delta_{1}$, as indicated. (c) Left axis: temperature dependencies of $\Delta_{1}$(T) for all five core-shell samples (left axis, $\square$=Co/CoO$\#$1, $\circ$=Co/CoO$\#$2, $\diamond$=Co/CoO$\#3$, $\bigtriangleup$=Co/CoO$\#4$, and $\triangledown$=Co/CoO$\#$5). Solid line is a mean field fit, showing that the inelastic scattering appears below the N\'{e}el temperature T$_{N}$=235 K, where the integrated intensities of the broad and narrow quasielastic scattering, integrated over wave vectors and normalized to their values at 325 K, vanish (right axis, broad quasielastic: $\blacksquare$=Co/CoO$\#$3, $\bullet$=Co/CoO$\#$4, narrow quasielastic: $\otimes$=Co/CoO$\#$3, $\oplus$=Co/CoO$\#$4).(d) The 300 K wave vector integrated broad quasielastic I$_{Broad}$ intensity and 150 K inelastic I$_{IN}$ intensity per nanoparticle are proportional for all five Co/CoO core shell samples. (e) The 150 K integrated inelastic intensity per particle and per nm$^3$ of CoO does not scale with the CoO mass. (f) The 150 K integrated inelastic intensity per nanoparticle I$_{IN}$ ($\Box$, left axis) and the 300 K broad quasielastic intensity I$_{Broad}$ ($\bigcirc$, right axis) are inversely proportional to the volume of the Co core V$_{Co}$. The dashed line indicates the background intensity I$_T$ of the thermally excited CoO spin waves. \label{data_fits_and_parameters}}
\end{figure*}

 I(q,E) is plotted at different temperatures in Fig.~\ref{overview}c for a wave vector cut which is 0.28 $\AA^{-1}$ wide, centered at q=1.40 $\AA^{-1}$. The indicated fits demonstrate that for T$\geq$ T$_{B}$, I(E) is well described as the sum  of two Lorentzian functions centered at E=0, but with very different widths.  Most of the scattered intensity is associated with the broader Lorentzian, I$_{Broad}$(q,E), and we will return to this component below. The Lorentzian I$_{Narrow}$(q,E) with the smaller linewidth $\Gamma_{Narrow}$ describes the superparamagnetic fluctuations of the Co core, where the core magnetization spontaneously reorients between two easy directions, separated by an energy barrier $\epsilon$(V$_{Co}$) = K(V$_{Co}$)V$_{Co}$, where V$_{Co}$ is the Co core volume and K is the volume dependent anisotropy constant. Fig.~\ref{data_fits_and_parameters}a shows that $\Gamma_{Narrow}$(T) decreases by $\approx$ 50$\%$ between 325 K and T$_{N}$, where it approaches the experimental energy resolution. The time scale $\tau \sim$ $\hbar$/$\Gamma_{Narrow}$ of these superparamagnetic fluctuations is governed by an Arrhenius-like temperature dependence $\tau=\tau_0exp(\epsilon/k_BT)$ \cite{wohlfarth1980}. By fitting this expression to the measured $\Gamma_{Narrow}$(T) (Fig.~\ref{data_fits_and_parameters}a, left panel), we extracted values for $\epsilon$(V$_{Co}$) for the Co/CoO$\#$3,4 samples of 62$\pm$6 and 61$\pm$3 meV, in acceptable agreement with the value of 78 meV reported previously for bare Co nanoparticles of a similar size.~\cite{chen1995} We were not able to resolve this narrow quasielastic scattering in samples Co/CoO$\#$1 and Co/CoO$\#$2, and we conclude that here the considerably larger Co cores have much longer reversal times $\tau$, leading to linewidths which are smaller than the 0.27 meV energy resolution of our experiment. The narrow Lorentzian component of the scattering found at high temperatures represents reversals of the Co cores which slow continuously with the approach to exchange blocking at T$_{B}\approx$200 K, as has previously been reported in neutron scattering studies of single component ferromagnetic, ferrimagnetic and antiferromagnetic  nanoparticles.\cite{mathieu1998b,lefmann1999,lefmann2001,klausen2004,schultheiss2008}

Figs.~\ref{overview}b,c show that inelastic neutron scattering is dominant once the quasielastic scattering is suppressed below T$_{B}$. We fit the scattered intensity below $\approx$ 200 K with a sum of four inelastic Lorentzian functions, centered at  $\pm$ E$_{1}$(q) and $\pm$ E$_{2}$(q). Fig.~\ref{data_fits_and_parameters}b shows that these excitations are weakly dispersing, with q=0 spin gaps $\Delta_{1,2}$(T). This scattering corresponds to bulk-like spin waves in the CoO shell. First, Fig.~\ref{data_fits_and_parameters}c shows that $\Delta_{1}$(T) has the appearance of an order parameter, dropping from a maximum value of ~2.65 meV at the lowest temperatures to zero near the common N\'{e}el temperature of the CoO shells T$_{N}$=235 K. Second, Fig.~\ref{data_fits_and_parameters}c shows that the magnitudes of the q-integrated $\Delta_{1}$ are identical in all five samples, despite their very different core and shell dimensions, suggesting that this excitation cannot be a surface or interface mode.  Fig.~\ref{overview}d compares the scattered intensity I(q,E) at 150 K for core-shell nanoparticles Co/CoO$\#$4, fully oxidized CoO nanoparticles, and a powder made by crushing a single crystal of CoO, each normalized by the total CoO mass. The inelastic scattering is strongest in the nanoparticle samples, particularly in the core-shell particles, and vanishingly weak in the bulk CoO powder. Nonetheless, the fits to I(q,E) indicated in Fig.~\ref{overview}d show that all three samples have the same inelastic excitations at E$_{1}$(q), albeit with different intensities and linewidths, confirming that these excitations are intrinsic to CoO. Scattering experiments using thermal neutrons were the first to identify spin waves with energies $\gtrsim$ 25 meV in a single crystal of CoO.~\cite{sakurai1968} A subsequent calculation of the effects of exchange coupling on the crystal field and spin-orbit split Co$^{2+}$ atomic state explained the weak dispersion of the bulk CoO spin waves, while predicting additional lower energy states, including one with a zone center energy of $\approx$ 3 meV,  which were subsequently observed in optical measurements.\cite{daniel1969} The dispersion predicted from these model parameters compares well to the ones found in our samples (Fig.~\ref{data_fits_and_parameters}b), with only a small decrease in the q=0 gap $\Delta_{1}$ in the nanoparticle system. For all these reasons, we conclude that we have observed bulk-like antiferromagnetic spin waves in both fully oxidized CoO nanoparticles, and in the CoO shells of composite Co/CoO core-shell nanoparticles. It is remarkable that bulk-like antiferromagnetic spin waves are robust in CoO shells which are only 4-10 unit cells thick, and that their dispersion is so similar to that of bulk CoO. The relative weakness of the exchange coupling relative to crystal field and spin orbit energies in CoO is responsible for the minimal dispersion of the spin waves, and we suggest that this inherently local character of the spin waves makes them more resistant to finite size effects than the strongly dispersing magnetic excitations found in metallic magnets, where these effects can be very large.~\cite{wesselinowa2007} In any case, these are the first direct observations by neutron scattering of dispersing spin waves with non-zero wave vectors reported in any nanoparticle system, ferromagnetic or antiferromagnetic.

We can now return to the broad quasielastic scattering I$_{Broad}$ found at T$\geq$T$_{B}$. Fig.~\ref{data_fits_and_parameters}d shows that in all five samples the intensity of this scattering is proportional to that of the inelastic spin wave scattering, confirming that both signals originate in the CoO shells. Given the very weak q-dependence of its linewidth and intensity, we ascribe I$_{Broad}$ to  paramagnetic fluctuations of individual moments in the CoO shells. The temperature dependence of the total scattering associated with I$_{Broad}$, normalized to its value at 325 K, is compared to those of the inelastic and narrow quasielastic signals in Fig.~\ref{data_fits_and_parameters}c. Consistent with the sequence outlined in Fig.~\ref{overview}b, we see that the quasielastic scattering I$_{Broad}$ is also suppressed when the core reversal is thermodynamically blocked as T$\rightarrow$T$_{B}$. Surprisingly, this suppression of I$_{Broad}$ is not accompanied by a slowing of the associated fluctuations. The right panel of Fig. ~\ref{data_fits_and_parameters}a shows instead that the linewidth of these paramagnetic fluctuations is almost temperature independent, and if anything \it increases \rm slightly with decreasing temperature. We will show that this surprising result is a direct consequence of the exchange coupling between the core and shell.

Our measurements find that the inelastic scattering depends on the CoO volume V$_{CoO}$ in an unexpected way. Fig.~\ref{data_fits_and_parameters}e reports that the integrated inelastic intensity at 150 K for the core - shell particles is not proportional to the CoO mass present, and that there is substantially more inelastic scattering present in the core - shell nanoparticles than in fully oxidized CoO nanoparticles or the bulk CoO powder.  Fig.~\ref{data_fits_and_parameters}f shows that the q- and E- integrated spin wave intensity has two components, I$_{IN}$=I$_{EX}$+I$_{T}$. The first, I$_{EX}$, increases linearly with 1/V$_{Co}$, demonstrating that the population of CoO spin waves is largest in core-shell particles where the cores are the smallest and where their dynamics are consequently the fastest. We take this as an overt signature of core-shell dynamical coupling. Consequently, this effect vanishes with the core dynamics as 1/V$_{Co}\rightarrow$0, and the vertical intercept represents the thermal population of spin waves I$_{T}$ which would normally be present in bare CoO nanoparticles of the same size. Fig.~\ref{data_fits_and_parameters}f shows that the presence of the Co core enhances the CoO spin wave population by as much as a factor of $\approx$2 in the sample with the smallest cores.

Since the core and shell share a common interface, it is impossible to modulate the core moments without simultaneously affecting the shell moments,~\cite{livesey2007} an effect which is only amplified in systems with high quality interfaces. For temperatures between T$_{N}$ and T$_{B}$, core-shell coupling strengthens, and leads to an enhanced population of shorter-lived paramagnetic fluctuations (Fig. 2a), overwhelming the critical slowing which would otherwise be expected for the ordering shells. For T$\leq$T$_{B}$, the dominant magnetic excitations of the core are the q=0 spin waves,~\cite{cehovin2003,morup2005} where core moments precess uniformly around the effective field direction H$_{eff}$, determined by both the exchange H$_{EB}$ and anisotropy K(V$_{Co}$)V$_{Co}$ fields. We propose that the decay of this mode, whose characteristic energy is $\omega_o\propto$MH$_{eff}$, generates additional Co spin waves with q$\neq$0 which are transmitted, at least in part, to the shell. In this scenario, the number of spin waves present in the CoO shell for T$\leq$T$_{B}$, beyond the thermal population, should be inversely proportional to the damping time for the core precessional mode.

\begin{figure*}
\includegraphics[width=9cm]{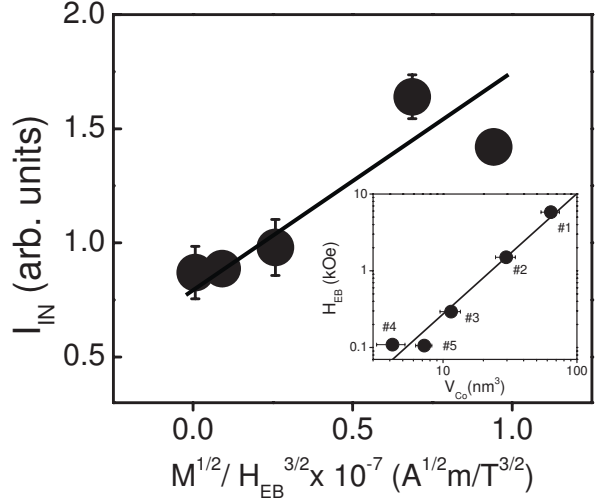}
\caption{\textbf{Lifetime of the Co Uniform Spin Waves} The wave vector integrated 150 K inelastic intensity per nanoparticle is proportional to the decay rate of the Co q=0 spin waves, estimated as 1/$\tau_{imp}\propto\sqrt{M}$/H$_{EB}^{3/2}$. Inset: the 30 K exchange bias field H$_{EB}$ depends approximately linearly on the Co core volume V$_{Co}$ for our neutron scattering samples Co/CoO\#1-5.\label{decay}}
\end{figure*}

The lifetime of the q=0 Co spin wave is limited by its decay via scattering into spin wave modes with q$\neq$0, and not by the spin-lattice relaxation time, which is known to be much longer.~\cite{kittel1953,vaterlaus1991,beaurepaire1996}. Antiferromagnetic exchange coupling shifts the precession mode energy $\omega_{0}$, and leads to enhanced and temperature dependent linewidths.~\cite{stamps1997,ercole2000}  In thin films, the q=0 mode decays into a pair of magnons with $\pm$q via scattering from random fields at the ferromagnetic/antiferromagnetic interface,~\cite{mathieu1998,arias1999,rezende2001} at a rate which is perhaps proportional to the exchange bias field itself.~\cite{weber2005}  Since the interfaces in our nanoparticles have near-epitaxial quality,~\cite{inderhees2008} it is more likely that impurity scattering is the primary decay mechanism for q=0 bulk spin waves in thermal equilibrium.~\cite{kittel1953,clogston1956a,clogston1956b} The rate for this process,~\cite{clogston1956a} 1/$\tau_{imp}\propto\sqrt{M}$/H$_{eff}^{3/2}$ is governed by the Co moment M and the effective field H$_{eff}$, which we take as H$_{EB}$.~\cite{jensen2001} We have plotted the 150 K inelastic intensity for each of the five neutron scattering samples as a function of $\sqrt{M}$/H$_{eff}^{3/2}$ in Fig.~\ref{decay}. We find a very reasonable linear relationship, confirming that impurity scattering limits the spin wave lifetime in our exchange biased nanoparticles, while validating our initial proposal that increasing the population of spin waves in the core results in a proportional increase in spin waves in the shell. Comparing Fig.~\ref{decay} and Fig. 2f implies that $\sqrt{M}$/H$_{eff}^{3/2}$ is proportional to 1/V$_{Co}$. Assuming that M$\propto$V$_{Co}$, then in turn H$_{EB}\propto V_{Co}$. The inset of Fig.~\ref{decay} confirms this relationship, in accord with previous reports.~\cite{nogues2005}

\begin{figure*}
\includegraphics[width=9cm]{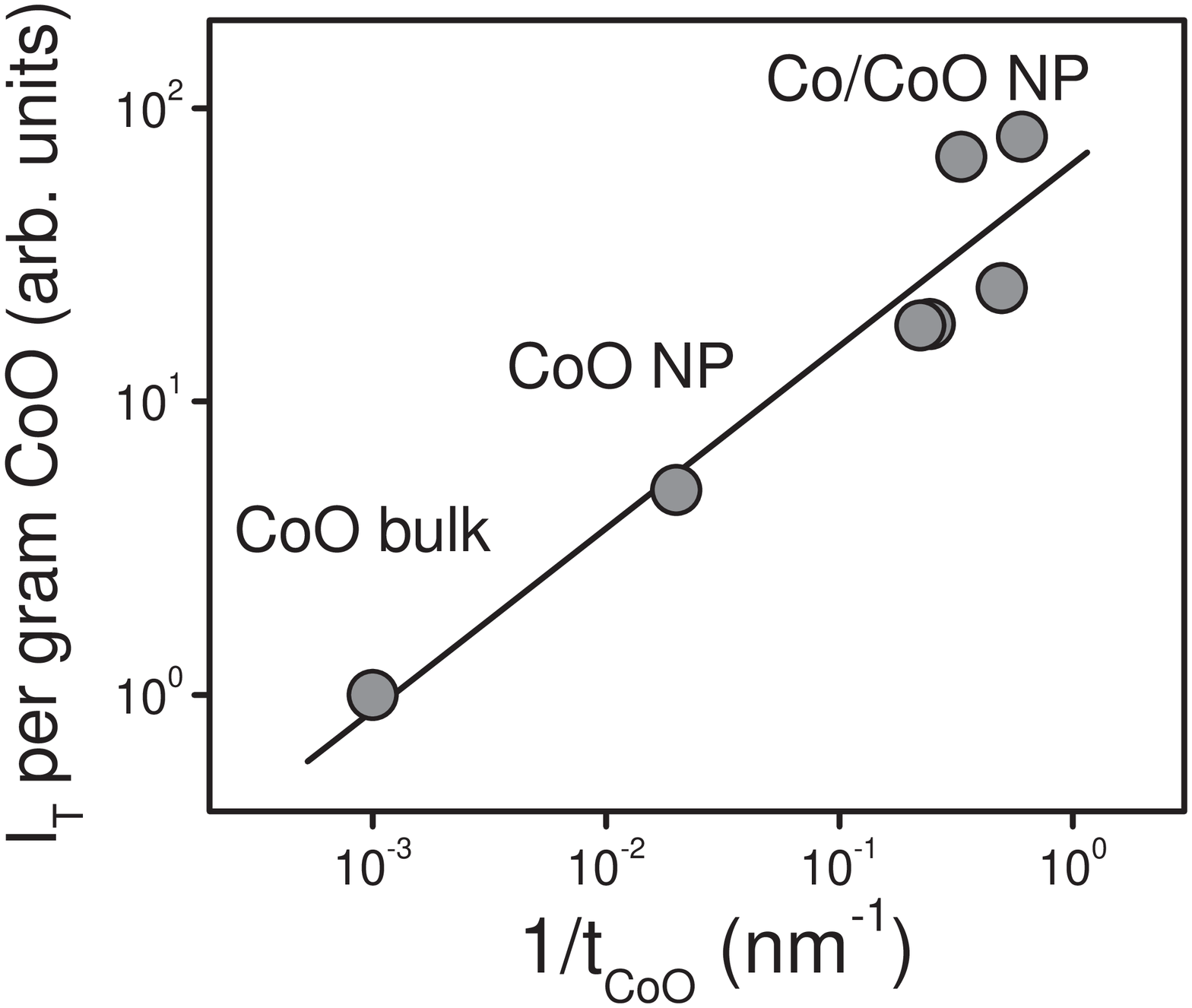}
\caption{\textbf{Finite size effects of the  CoO Spin Waves} The part of the 150 K spin wave population in the core-shell nanoparticles which is caused by core reorientation I$_{EX}$ is subtracted from the integrated spin wave intensities
 to isolate the thermally excited spin wave intensity I$_{T}$, which is compared here to those measured in CoO nanoparticles and a fine-grained powder, prepared by crushing a single crystal of CoO. These intensities I$_{T}$, normalized by the total mass of CoO,  are plotted as a function of the inverse of the CoO length scale t$_{CoO}$, which is
 the shell thickness of the core-shell nanoparticles, the CoO nanoparticle radius, and the average grain radius for the powder. The solid line demonstrates that these data are well described by the power law
 I$_{T}\propto$t$_{CoO}^{0.6\pm0.1}$. \label{thermal}}
\end{figure*}

Finite size effects impact the CoO spin waves in dramatic ways. Fig.~\ref{overview}d shows that the CoO spin wave scattering for T$\leq$T$_{B}$ is present at levels far in excess of those found in CoO nanoparticles or in bulk CoO. Part of this effect arises from coupling to the core dynamics. In addition, we will show that there is a dramatic redistribution of the thermally excited spectral weight as the CoO system is reduced in size.  Fig.~\ref{thermal} compares the q- and E- integrated inelastic intensity at 150 K for three different forms of CoO: bulk powder, CoO nanoparticles, and nanoparticle shells, where in the latter case we have plotted only the thermal part of the spin wave population I$_{T}$ as a function of the inverse system size. Within our experimental window in q and E, it is clear that much of the spectral weight which is concentrated in antiferromagnetic Bragg peaks in bulk CoO migrates to spin wave excitations as the system size becomes smaller. Even though our measurements  show that the spin wave dispersion of CoO is itself not significantly modified by finite size effects, Fig.~\ref{overview}a shows that the spin wave population is strongly shifted towards larger wave vectors and smaller wavelengths in nanoscaled CoO, relative to bulk CoO where little q-dependence is observed beyond the magnetic form factor.~\cite{sakurai1968,daniel1969} This dynamical broadening and shifting of spin wave intensity to larger wave vectors was predicted in numerical studies of very small ferromagnetic particles,~\cite{hendriksen1993} and it is interesting and perhaps surprising that this effect is apparently still robust in our CoO shells which have as many as 10$^{4}$ - 10$^{5}$ Co moments.

In summary, we have reported here the first observation of dispersing spin waves in a nanoparticle system, made on Co core/CoO shell nanoparticles using inelastic neutron scattering measurements. The spin waves originate in the CoO shells of the nanoparticles, and their dispersion is very similar to that of bulk CoO. The spin wave population is largest in core-shell nanoparticles where the core is smallest and most quickly reorienting, indicating that the dynamical coupling between the core and shell provides a new pathway for dissipating the energy of core reversal and reorientation via a generation of auxiliary spin waves in the shell.  It is possible that this mechanism is responsible for the enhanced switching speed which has been reported in some exchange biased systems,~\cite{jensen2001} confirming the paradoxical conclusion that the exchange bias effect improves both the stability of the moments of small ferromagnetic particles as well as the rate at which they can be intentionally switched. The CoO spin waves are strengthened by finite size effects, especially at the shortest wavelengths. Even in the N\'{e}el ordered and dynamically blocked state, the magnetization of a nanoparticle is modulated on shorter length scales and persists for shorter times than its bulk analog. This relative primacy of the spin waves and their dispersion must be included in any realistic picture of the equilibrium properties and the switching dynamics of magnetic nanoparticles.


\paragraph *{\bf{Acknowledgements}}
\vspace{2 mm} We are grateful to L. Yang and A. Kou for helping
with SAXS measurements and data analysis, and to G. Strycker for
assistance with data analysis. We acknowledge S. Kline, K. Dahmen,
and M. Weissman for useful discussions. This work was carried out under the
auspices of the U.S. Department of Energy, Office of Basic Energy
Sciences at both the University of Michigan and at Brookhaven
National Laboratories, the latter under Contract No.
DE-AC02-98CH1886. Use of the National Synchrotron Light Source,
Brookhaven National Laboratory, was supported by the U.S.
Department of Energy, Office of Science, Office of Basic Energy
Sciences, under Contract No. DE-AC02-98CH10886. Part of this
research was carried out at the Center for Functional
Nanomaterials, Brookhaven National Laboratory, which is supported
by the U.S. Department of Energy, Office of Basic Energy Sciences,
under Contract No. DE-AC02-98CH10886. We acknowledge the support
of the National Institute of Standards and Technology, U.S.
Department of Commerce, in providing some of the neutron research
facilities used in this work. Work at NIST is supported in part by the National Science Foundation under Agreement No. DMR-0454672.
Research at Oak Ridge National
Laboratory's Spallation Neutron Source was sponsored by the
Scientific User Facilities Division, Office of Basic Energy
Sciences, U. S. Department of Energy. The work of X.T and W. D. is partially
supported by the University of New Hampshire.
\paragraph *{\bf{Author contributions}}
All authors discussed the results and commented on the manuscript. M.C.A., S.E.I, and M.F conceived and designed the experiments and analyzed the data. M. F. and M.C.A co-wrote the manuscript.
S.E.I, X.T, W.D, W.H, J. W., and Z. X. provided the samples used in the experiments. A.P, J.N., M. H., Y.Q., and C.B. provided support for neutron scattering experiments, while Y.Y. and L.Z.
provided technical assistance with other measurements.
\paragraph *{\bf{Competing financial interests}}
The authors declare that they have no competing financial interests.
\paragraph *{\bf{Methods}}
\vspace{2 mm}
\paragraph *{\bf{High-resolution transmission electron microscopy (HRTEM)}}
\vspace{2 mm} HRTEM studies were carried out at Center for Functional Nanomaterials at Brookhaven National Laboratory and at the Electron Microbeam Analysis Laboratory (EMAL). Specimens were prepared by dispersing surfactant-free nanoparticles in chloroform (~1 mg/mL) and then depositing them on carbon coated-copper grids by drop-casting. HRTEM at BNL was performed using a field-emission JEM 2100F and at Michigan using a JEOL 3011 microscope, both equipped with energy dispersive X-ray spectrometry (EDS). Image acquisition and analysis were performed using a Gatan Digital Micrograph.
\paragraph *{\bf{Magnetization measurements}}
\vspace{2 mm}dc magnetization measurements of precipitated nanoparticle powders suspended in paraffin were performed using the Quantum Designs MPMS-XL magnetometer. To estimate the exchange bias field H$_{EB}$, the samples were cooled from 300 K to 10 K in a magnetic field of 5 T. The M (H) hysteresis loop was measured in the field range from -5 T to 5 T, and we have subtracted a linear contribution from M(H) prior to determining H$_{EB}$.
\paragraph *{\bf{Small-angle x-ray scattering}}
\vspace{2 mm} Small-angle x-ray scattering experiments were performed at beamline X21 at the National Synchrotron Light Source (NSLS) at Brookhaven National Laboratory,  with x-ray wavelength $\lambda = 1.23$\AA.
A dilute suspension of Co/CoO nanoparticles in toluene was contained in quartz capillaries, and the data were corrected for the background scattering and beam attenuation.
\paragraph *{\bf{Inelastic neutron scattering}}
\vspace{2 mm} Inelastic neutron measurements were carried out at the NIST Center for Neutron Research using the Disk Chopper Spectrometer for samples Co/CoO$\#$1-4 using an incident neutron wave length of 3.50 \AA, and on samples Co/CoO$\#$4-5, fully oxidized CoO nanoparticles, and a bulk powder of CoO using the Cold Neutron Chopper Spectrometer (CNCS) at the Spallation Neutron Source using a neutron wave length of 2.34 \AA. Each sample consisted of several grams  of precipitated nanoparticle powders from which the surfactants have been removed, which were subsequently wrapped in Al foil and then sealed in double walled Al sample cans. The measured scattering intensities were normalized to the signal from a vanadium sample and the background from the empty sample holder, while negligible, was subtracted. Data reduction and analysis were performed using the DGSreduction and DAVE software packages. \cite{dave}.

Identification of commercial equipment or products in the text is not intended to imply any recommendation or endorsement by the National Institute of Standards and Technology.

\vspace{2 mm}

\paragraph *{\bf{References}}

\end{document}